# Valley kink states and topological channel intersections in substrate-integrated photonic circuitry

Li Zhang[1,2,#], Yihao Yang[1,2,3,4, #,*], Mengjia He[1,2, #], Hai-Xiao Wang[5, #], Zhaoju Yang[3], Erping Li[1], Fei Gao[1,2], Baile Zhang[3,4], Ranjan Singh[3,4], Jian-Hua Jiang[5,*], and Hongsheng Chen[1,2,*]

[1]Key Lab. of Advanced Micro/Nano Electronic Devices & Smart Systems of Zhejiang, College of Information Science and Electronic Engineering, Zhejiang University, Hangzhou 310027, China.

[2]State Key Laboratory of Modern Optical Instrumentation, and The Electromagnetics Academy at Zhejiang University, Zhejiang University, Hangzhou 310027, China.

[3]Division of Physics and Applied Physics, School of Physical and Mathematical Sciences, Nanyang Technological University, 21 Nanyang Link, Singapore 637371, Singapore.

[4]Centre for Disruptive Photonic Technologies, The Photonics Institute, Nanyang Technological University, 50 Nanyang Avenue, Singapore 639798, Singapore.

[5]Shool of Physical Science and Technology, & Collaborative Innovation Center of Suzhou Nano Science and Technology, Soochow University, Suzhou 215006, China.

# These authors contributed equally to this work.

[*] *yang.yihao@ntu.edu.sg* (Yihao Yang); *jianhuajiang@suda.edu.cn* (Jian-Hua Jiang); *hansomchen@zju.edu.cn* (Hongsheng Chen).

## ABSTRACT

Valley degrees of freedom, providing a novel way to increase capacity and efficiency of information processing, have become an important instrument for photonics. Experimental studies on photonic topological valley kink states at interfaces between regions with opposite valley-Chern numbers have attracted much attention, however, were restricted to zigzag-type interfaces, largely limiting their applications such as geometry-dependent topological channel intersections. Here, we experimentally demonstrate and manipulate valley kink states at generic interfaces in subwavelength substrate-integrated photonic circuitry. We verify the robustness of the kink states by measuring transmissions of the kink states through twisted interfaces and interfaces with disorders. Based on the valley kink states at generic interfaces, we realize several topological channel intersections where photonic transport paths are related to geometries of the intersections. In comparison to those in previous works, the present valley photonic crystals have subwavelength thicknesses and excellent self-consistent electrical shielding, which are perfectly compatible with conventional substrate-integrated photonic circuitry. Our work opens a door to manipulate photonic valley pseudo-spins in substrate-integrated circuitry with robustness, easy access, and lightweight.

## INTRODUCTION

Topological photonics, which studies topological states of photons, has attracted much attention in the past decade[1-3]. Researches on topological photonics have shown emergence of various edge/surface states in topological photonic systems and their unique phenomenology, including chiral/helical edge propagation[4-12], pseudospin-momentum locking[5,13], topologically-protected refraction[14], and Fermi-arc surface states[15-17]. These properties have led to unprecedented applications in photonics such as backscattering-immune waveguides[7], robust delay lines[18], topological waveguide splitters[5, 19], and topological lasers[5, 19-22].

Recent studies have revealed a class of topological interface states that emerge at interfaces between regions with opposite valley-Chern numbers, also known as topological valley kink states[14, 19, 23-34]. The topological valley kink states have been proven to exist at generic interfaces, such as zigzag, armchair, and any combination of zigzag and armchair interfaces, in electronic[26, 28, 30-32], magnonic[25], and elastic[27] systems. The valley kink states are robust against some disorders due to the bulk-edge correspondence, with their topological numbers characterized by the difference of the valley-Chern numbers across the interface[25, 31]. Such valley transport along generic interfaces can also have many potential applications in photonics, such as topological Mach-Zehnder interferometers[25], high-efficiency energy transport channels[26], and topological channel intersections[27, 30, 32]. However, previous experimental realizations of the photonic topological valley kink states were restricted to zigzag interfaces[14, 19, 24, 33-35], largely limiting the above applications.

In this work, we demonstrate experimentally valley-polarized topological kink states at generic interfaces in subwavelength substrate-integrated photonic circuitry. We verify the robustness of the topological kink states by measuring transmissions of kink states through twisted interfaces and interfaces with disorders. Based on the valley kink states at generic interfaces, we demonstrate topological channel intersections where photonic transport paths are determined by geometries of the intersections. Besides, in comparison to the previous valley photonic crystals (VPCs)[14, 19, 35], the proposed VPCs have subwavelength thicknesses and excellent self-consistent electrical shielding, which are perfectly compatible with conventional substrate-integrated photonic circuits[36]. The manipulation of valley kink states at will in substrate-integrated photonic circuitry paves the way for future valley photonics.

## RESULTS

The designed VPC consists of a hexagonal-lattice of triangular scatterers placed in a copper parallel plate waveguide, which is loaded with a dielectric material with relative permittivity 2.4 (see Fig. 1b). The lattice constant is $a$ =10.4 mm, and the conductivity of the copper is $5.7\times10^7$ S/m. A triangular scatterer consists of three via holes with diameter $d_0$=0.8 mm and height $h$ =3.2 mm, as well as a triangle copper patch with side length $l$ =5.2 mm at the top. There is a gap with $g_0$=0.25 mm between the scatterers and the top metal cladding layer. The gap size $g_0$ plays an important role in controlling the bandgap size (see Supplementary Note 1). The rotation angle $\theta$ is defined in Fig. 1b. In our experiments, the VPCs consist of two printed circuit boards (PCBs). The copper triangle patch and via holes are printed on the bottom PCB. The upper side of the top PCB is covered with a thin layer of copper. We then attach the bottom PCB to the top PCB with an adhesive (see Fig. 1a). In comparison to the previous VPCs[14, 19, 35], the present structure shows the merits of a subwavelength thickness (only 0.245 $\lambda_l$ , where $\lambda_l$ is the wavelength in the loaded dielectric material) and excellent electromagnetic shielding, which are perfectly compatible with the standard substrate-integrated waveguide circuits. Besides, the loss tangent of the substrate used is around 0.001, and the copper behaves as a perfect electric conductor at low frequencies, therefore, the absorptive loss is low in our case, which has been verified in our numerical results (see Supplementary Note 2).

When the rotation angle is $\theta$=0°, there are a pair of Dirac points at K and K' points in the first Brillouin zone at a frequency of 14.39 GHz (see Fig. 1c). With a finite rotation angle $\theta$, the symmetry of the photonic crystal is reduced from $C_{3V}$ to $C_3$. Consequently, the pair of Dirac points are gapped out, resulting in a photonic bandgap controlled by the rotation angle $\theta$. The photonic band structure for $\theta$ =30° is shown in Fig. 1c: a bandgap appears from 13.43 GHz to 15.12 GHz. The field patterns of the upper and lower bands at K and K' points are shown in Fig. 1e, which demonstrates the Poynting vectors are related to photonic valley pseudo-spins. Note that all simulations in this work are performed in 3D using the realistic structures (see Supplementary Note 3).

The band topology around K valley can be described by an effective massive Dirac Hamiltonian[14, 19, 29, 33-35]. For instance, the Hamiltonian for the K valley is

$$H_K = v(q_x\sigma_x + q_y\sigma_y) + m\sigma_z \quad (1)$$

where $\sigma_z = \pm 1$ represent $\varphi_+$ and $\varphi_-$ states, respectively, which carry opposite valleys; $\vec{q} = (q_x, q_y)$ is wavevector derivation from K point; and $m$ is an effective mass term determined by the bandgap size. In parallel, the band topology around K' point is described by a different Hamiltonian, $H_{K'} = -H_K$, as guaranteed by the time-reversal symmetry preserved in our system. These Hamiltonians indicate the nontrivial Berry curvatures around the K and K' point. Integrating the Berry curvature around K (K') valley gives a half-integer valley-Chern number $C_K = \mathrm{sgn(m)}/2$ [14, 19, 29, 33-35]. First-principle numerical calculations of the Berry curvatures near the K and K' valleys for our practical VPCs are presented in Supplementary Note 4.

The rotation angle $\theta$ controls the Dirac mass terms and the corresponding valley-Chern number (see Fig. 1d). The photonic band structure experiences phase transitions when $\theta$ crosses an integer time of 60°. Across an interface that consisting of two regions with opposite valley-Chern numbers, the difference of the valley-Chern numbers is $\Delta C_K = \mathrm{sgn(m)} = \pm 1$ for K and K' valleys, respectively, giving rise to valley-polarized topological kink states according to the bulk-edge correspondence[14, 25, 28, 29, 31].

To verify the above statements, we numerically calculate two elementary types of interfaces. The first type is the previously-studied zigzag interfaces[14, 19, 33-35] (see Figs. 2a-2b). The kink states at two different zigzag interfaces exhibit different spatial symmetries: mirror-symmetric (Fig. 2a) and anti-symmetric (Fig. 2b). The photonic band structure in Fig. 2c indicates that the propagating directions of the two topological kink states bounded to different valleys are exactly opposite, exhibiting "valley-locked" chirality.

The second type is armchair interfaces (Figs. 2d and 2e). The pair of kink states are bounded to different valleys, and they propagate in opposite directions along the interface[25, 31]. We note that a tiny, avoided crossing bandgap appears at the crossing point (i.e., $k$=0) between two counter-propagating kink states[30,31], similar to the tiny bandgaps of the edge states in photonic topological insulators with $C_6$ crystalline symmetry[4, 11, 12]. Besides, we also perform simulations of two additional interfaces different from the zigzag and armchair interfaces, each of which shows a pair of kink states bounded to the different valleys (Supplementary Note 5).

It is well-known that conventional photonic crystal waveguides suffer considerable reflection loss when electromagnetic waves propagate through sharp bending corners[37]. In contrast, our experimental and numerical results indicate that the valley kink states host robust transport: electromagnetic waves propagating along the topological interfaces can turn around the sharp bending corners with negligible backscattering (see Fig. 3). Such robust transport is observed in our VPCs for both the zigzag and armchair interfaces (see Figs. 3a, 3b, 3d, and 3e). We further demonstrate the valley kink states in a photonic circuitry with the interconnection between zigzag and armchair interfaces (see Figs. 3c and 3f). The measured transmissions for all three samples are quantitatively analyzed in Fig. 3g and 3h. High transmissions of valley transport are observed for all photonic circuitry, whereas the counterpart of bulk mode is ultralow. Note that as the gap of the kink state dispersion at the armchair interface is so small that we do not observe obvious dips in the transmission spectra for the armchair interface in the experiments. The electromagnetic field-intensity profiles of the kink states near the output terminal are also measured (See Supplementary Note 6 for details), showing the kink states are confined around the interface. Most significantly, within the bandgap, the transmissions through various bending interfaces are close to the counterpart through the straight interfaces, indicating suppressed back-scattering and robust transport of the valley kink states (see Fig. 3g). Besides, we would like to mention that in both simulations and experiments, all interface terminations have been truncated to the zigzag type to suppress the reflections (See Supplementary Note 7 for details). Also, the out couplings of the kink states from the VPCs to empty parallel waveguides for different types of terminations have been numerically studied. The results manifest that reflections are negligible at the zigzag terminations while considerable at the armchair terminations (See Supplementary Note 8 for details).

We further confirm the robust transport of topological kink state by introducing certain disorders into VPCs around the interfaces. As shown in Fig. 4, valley kink states are robust against the randomly placed structure disorders (by flipping the rotation angles of the triangular scatterers, see Figs. 4a and 4c). One can see that the transmissions of the valley transport are very close in the cases with and without disorders. The field-intensity profiles manifest that the kink states in the "bus" waveguide couples to the disorder-induced cavities, and then the trapped electromagnetic wave finally couples back to the "bus" waveguide with negligible backscattering (Figs. 4b and 4d). This is because in the disorder-induced cavities, fields of the modes belonging to different valleys vary between negative and positive and overlap, and the variations tend to cancel each other out[29].

Besides, we perform additional simulations to demonstrate the robustness of the valley kink states against different kinds of disorders (see Supplementary Note 9).

Finally, based on the valley kink states at generic interfaces, we realize photonic geometry-dependent topological channel intersections, where energy transport paths are related to geometries of the intersections[27, 30, 32]. We design three different cross-shaped channel intersections (see Figs. 5a-5c). The first cross-shaped channel intersection consists solely of zigzag interfaces (Fig. 5a), where the corresponding valley-polarized topological kink states are marked with red and blue arrows for different valleys. In the channels, electromagnetic waves travel only along the paths of the same valley degree of freedom. From our simulations (Figs. 5(d)-(e)), one can see that the valley kink states launched at port 1 (port 2) can only transport along path 2 and path 4 (path 1 and path 3) and are forbidden to travel along path 3 (path 4). We can understand the interesting phenomenon in the following way: when the kink state bounded to K valley is excited from port 1, it cannot couple to path 3 because the kink states at the two paths have different valley pseudo-spins. However, it can couple to path 2 and path 4, owing to the same valley pseudo-spins.

Experiments are also conducted to demonstrate the geometry-dependent topological channel intersections. In the experiments, a source antenna is placed at port 1 (port 2) of the sample, whereas probe antennas are placed at the other three output terminals to detect kink states bounded to K (K') valley. From Fig. 5(j), one can see that the transmissions S21 and S41 are much higher than S31 (see the upper panel), whereas the transmissions S12 and S32 are much higher than S42 (see the lower panel), within the photonic bandgap frequency window. Similar phenomena have also been observed for a channel intersection with solely armchair interfaces (Figs. 5b, 5f-g, and 5k) and that with both zigzag and armchair interfaces (Figs. 5c, 5h-i, and 5l) in our photonic crystal. These observations confirm that the valley degrees of freedom are bounded to the propagations of the topological kink states at each channel and the energy transport at the channel intersections are directly related to the geometries. Due to the impedance and mode mismatches between the excitation sources and the topological interfaces, the average measured transmission is lower than 3 dB. To further improve the transmission efficiency, one can design smooth bridges between the excitation source and the topological interfaces, similar to the spoof surface plasmon polariton couplers [38].

## CONCLUSION AND DISCUSSIONS

We thus experimentally demonstrate and manipulate topological valley kink states at generic interfaces in subwavelength substrate-integrated photonic circuitry. Such valley kink states are robust against some types of disorders and sharp bends, as we verified both experimentally and numerically. Based on the valley kink states at generic interfaces, we realize topological channel intersections where photonic transport paths are related to geometries of the intersections. Moreover, in comparison to the previous VPCs, our substrate-integrated VPCs show subwavelength thicknesses and excellent self-consistent electrical shielding, which are perfectly compatible with the conventional substrate-integrated waveguide circuitry. The present design provides a possibility for a complete topological waveguide circuitry, including passive, active, or nonplanar components and even antennas, to be fabricated and integrated on the same substrate using standard printed circuits board techniques. Our studies add new freedom, namely, topology, beyond the conventional freedoms, i.e., wave-vector, phase, and polarization, to the substrate-integrated circuitry, and may work as an alternative integration platform with low cost, easy access, and light weight.

## Supporting Information

Supporting Information is available from the Wiley Online Library or from the author.

## Acknowledgments

Work at Zhejiang University was sponsored by the National Natural Science Foundation of China under Grants No. 61625502, No. 61574127, and No. 61601408, the Top-Notch Young Talents Program of China, and the Innovation Joint Research Center for Cyber-Physical-Society System. Work at Soochow University is supported by the National Natural Science Foundation of China under Grant No. 11675116 and the Jiangsu distinguished professor funding. We thank Y. D. Chong at Nanyang Technological University for helpful discussions.

## Conflict of Interests

The authors declare no conflict of interests.

## REFERENCES

[1] L. Lu, J. D. Joannopoulos, M. Soljačić, *Nat. Photonics* **2014**, 8, 821.

[2] T. Ozawa, H. M. Price, A. Amo, N. Goldman, M. Hafezi, L. Lu, M. Rechtsman, D. Schuster, J. Simon, O. Zilberberg, *Rev. Mod. Phys.* **2018**, 80, 015006

[3] A. B. Khanikaev, G. Shvets, *Nat. Photonics* **2017**, 11, 763.

[4] S. Yves, R. Fleury, T. Berthelot, M. Fink, F. Lemoult, G. Lerosey, *Nat. Commun.* **2017**, 8, 16023.

[5] X. Cheng, C. Jouvaud, X. Ni, S. H. Mousavi, A. Z. Genack, A. B. Khanikaev, *Nat. Mater.* **2016**, 15, 542.

[6] A. B. Khanikaev, S. H. Mousavi, W. K. Tse, M. Kargarian, A. H. MacDonald, G. Shvets, *Nat. Mater.* **2013**, 12, 233.

[7] Z. Wang, Y. Chong, J. D. Joannopoulos, M. Soljacic, *Nature* **2009**, 461, 772.

[8] M. Hafezi, S. Mittal, J. Fan, A. Migdall, J. Taylor, *Nat. Photonics* **2013**, 7, 1001.

[9] T. Ma, A. B. Khanikaev, S. H. Mousavi, G. Shvets, *Phys. Rev. Lett.* **2015**, 114, 127401.

[10] Y. Poo, R. X. Wu, Z. Lin, Y. Yang, C. T. Chan, *Phys. Rev. Lett.* **2011**, 106, 093903.

[11] L. H. Wu, X. Hu, *Phys Rev. Lett.* **2015**, 114, 223901.

[12] S. Barik, A. Karasahin, C. Flower, T. Cai, H. Miyake, W. DeGottardi, M. Hafezi, E. Waks, *Science* **2018**, 359, 666.

[13] Y. Yang, Z. Gao, H. Xue, L. Zhang, M. He, Z. Yang, R. Singh, Y. Chong, B. Zhang, H. Chen, *Nature* **2019**, 565, 622.

[14] F. Gao, H. Xue, Z. Yang, K. Lai, Y. Yu, X. Lin, Y. Chong, G. Shvets, B. Zhang, *Nat. Phys.* **2017**, 14, 140.

[15] B. Yang, Q. Guo, B. Tremain, L. E. Barr, W. Gao, H. Liu, B. Béri, Y. Xiang, D. Fan, A. P. Hibbins, *Nat. Commun.* **2017**, 8, 97.

[16] J. Noh, S. Huang, D. Leykam, Y. D. Chong, K. P. Chen, Mikael C. Rechtsman, *Nat. Phys.* **2017**, 13, 611.

[17] B. Yang, Q. Guo, B. Tremain, R. Liu, L. E. Barr, Q. Yan, W. Gao, H. Liu, Y. Xiang, J. Chen, C. Fang, A. Hibbins, L. Lu, S. Zhang, *Science* **2018**, 359, 1013.

[18] M. Hafezi, E. A. Demler, M. D. Lukin, J. M. Taylor, *Nat. Phys.* **2011**, 7, 907.

[19] X. Wu, Y. Meng, J. Tian, Y. Huang, H. Xiang, D. Han, W. Wen, *Nat. Commun.* **2017**, 8, 1304.

[20] B. Bahari, A. Ndao, F. Vallini, A. El Amili, Y. Fainman, B. Kante, *Science* **2017**, 358, 636.

[21] M. A. Bandres, S. Wittek, G. Harari, M. Parto, J. Ren, M. Segev, D. N. Christodoulides, M. Khajavikhan, *Science* **2018**, 359, eaar4005.

[22] G. Harari, M. A. Bandres, Y. Lumer, M. C. Rechtsman, Y. D. Chong, M. Khajavikhan, D. N. Christodoulides, M. Segev, *Science* **2018**, 359, eaar4003.

[23] J. Lu, C. Qiu, L. Ye, X. Fan, M. Ke, F. Zhang, Z. Liu, *Nat. Phys.* **2016**, 13, 369.

[24] J. Noh, S. Huang, K. P. Chen, M. C. Rechtsman, *Phys. Rev. Lett.* **2018**, 120, 063902.

[25] Y. M. Li, J. Xiao, K. Chang, *Nano. Lett.* **2018**, 18, 3032.

[26] Z. Qiao, J. Jung, Q. Niu, A. H. Macdonald, *Nano. Lett.* **2011**, 11, 3453.
[27] M. Yan, J. Lu, F. Li, W. Deng, X. Huang, J. Ma, Z. Liu, *Nat. Mater.* **2018**, 17, 993.
[28] L. Ju, Z. Shi, N. Nair, Y. Lv, C. Jin, J. Velasco, Jr., C. Ojeda-Aristizabal, H. A. Bechtel, M. C. Martin, A. Zettl, J. Analytis, F. Wang, *Nature* **2015**, 520, 650.
[29] T. Ma, G. Shvets, *New J. Phys.* **2016**, 18, 025012.
[30] Z. Qiao, J. Jung, C. Lin, Y. Ren, A. H. MacDonald, Q. Niu, *Phys. Rev. Lett.* **2014**, 112, 206601.
[31] Y. Ren, Z. Qiao, Q. Niu, *Rep. Prog. Phys.* **2016**, 79, 066501.
[32] J. Li, R.-X. Zhang, Z. Yin, J. Zhang, K. Watanabe, T. Taniguchi, C. Liu, J. Zhu, *Science* **2018**, 362, 1149.
[33] X.-T. He, E.-T. Liang, J.-J. Yuan, H.-Y. Qiu, X.-D. Chen, F.-L. Zhao, J.-W. Dong, *Nat. Commun.* **2019**, 10, 872.
[34] M. I. Shalaev, W. Walasik, A. Tsukernik, Y. Xu, N. M. Litchinitser, *Nat. Nanotech.* **2019**, 14, 31.
[35] Z. Gao, Z. Yang, F. Gao, H. Xue, Y. Yang, J. Dong, B. Zhang, *Phys. Rev. B* **2017**, 96, 201402.
[36] M. Bozzi, A. Georgiadis, K. Wu, *IET Microw. Anten. P.* **2011**, 5, 909.
[37] A. Mekis, J. Chen, I. Kurland, S. Fan, P. R. Villeneuve, J. Joannopoulos, *Phys. Rev. Lett.* **1996**, 77, 3787.
[38] H. F. Ma, X. Shen, Q. Cheng, W. X. Jiang, T. J. Cui, *Laser Photonics Rev.* **2014**, 8, 146.

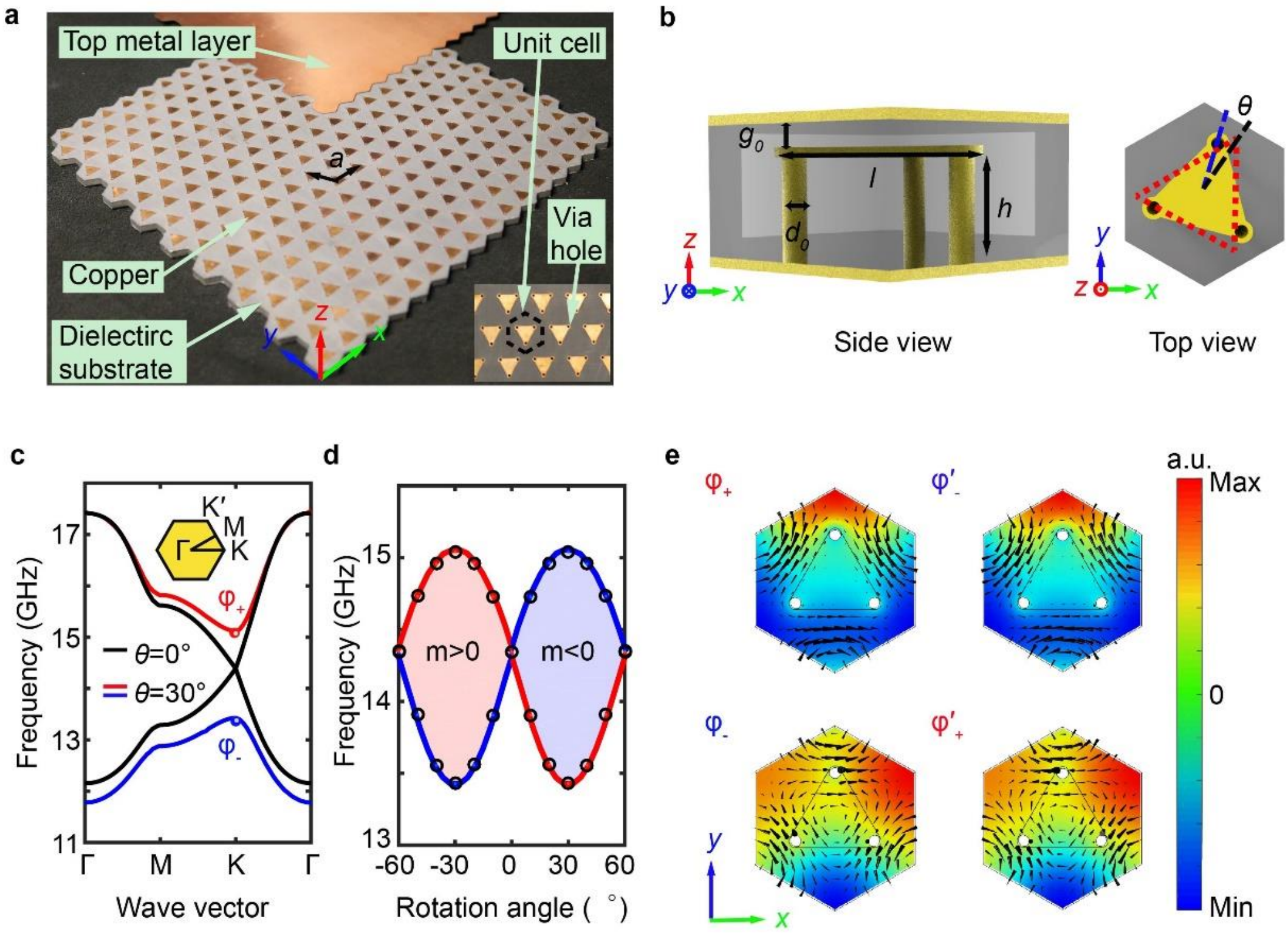

**Figure 1. Substrate-integrated VPCs with subwavelength thicknesses. a,** Perspective-view photograph of the experimental sample. **b**, Side and top views of a unit cell with a rotation angle of 15°. The side view shows the details of the sample, and the top view shows the sample without the top PCB. In the top view, the red dashed triangle indicates the triangle scatterer with a rotation angle of 0°. The angle $\theta$ between the black and blue dashed lines represents the rotation angle. The yellow area represents the PEC structure, and the grey region represents the dielectric substrate. **c**, Band structures of the VPC with a rotation angle of 0° (black curve) and 30° (colored curves), respectively. **d**, Band edges and effective Dirac mass term at the K point as a function of the rotation angle $\theta$. **e**, Field patterns of four eigenstates at band edges: K point ( $\varphi_+$ and $\varphi_-$ ) and K' point ( $\varphi'_+$ and $\varphi'_-$ ). Black arrows represent Poynting vector distributions for each state.

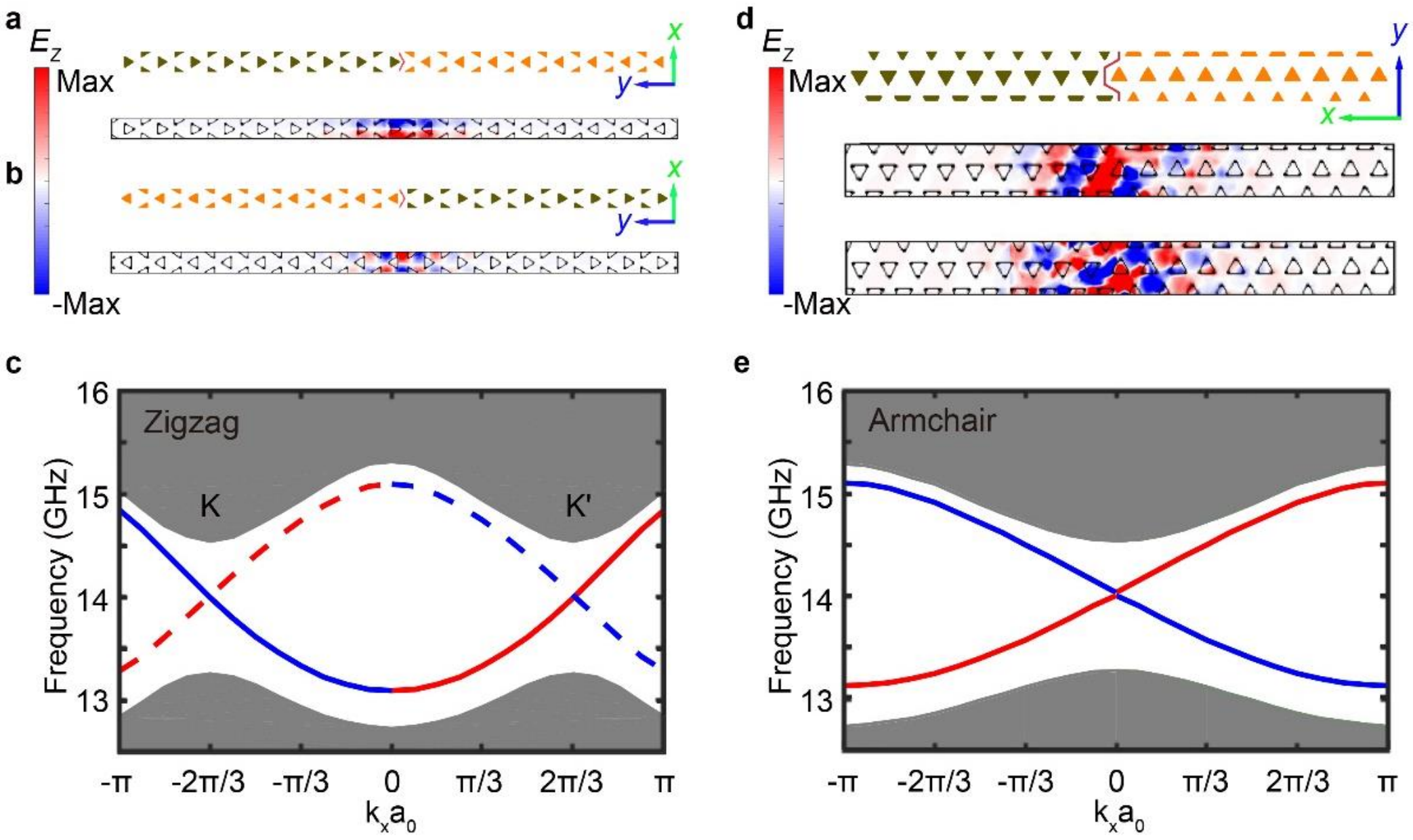

**Figure 2. Zigzag and armchair interfaces and valley-polarized topological kink states. a, b**, Zigzag interfaces formed by VPCs with opposite rotation angles. The color profiles represent the electric field distributions of the valley kink states at the K' point. **c**, Band structures of zigzag interfaces in **a** and **b**. Here, gray regions, dashed curve, and solid curve represent projected bulk bands and dispersions of interfaces in **a** and **b**, respectively. The red and blue curves indicate forward- and backward-propagating kink states, respectively. **d**, Armchair interfaces formed by VPCs with opposite rotation angles. The color profiles represent electric field distributions of topological kink states at K (upper) and K' (lower) valleys. **e**, Band structures of the armchair interface in **d**. There is a tiny avoided crossing bandgap between two kink states (almost negligible in the present scale). Here, gray regions denote projected bulk bands, whereas red (blue) curves represent kink states of positive (negative) group velocities.

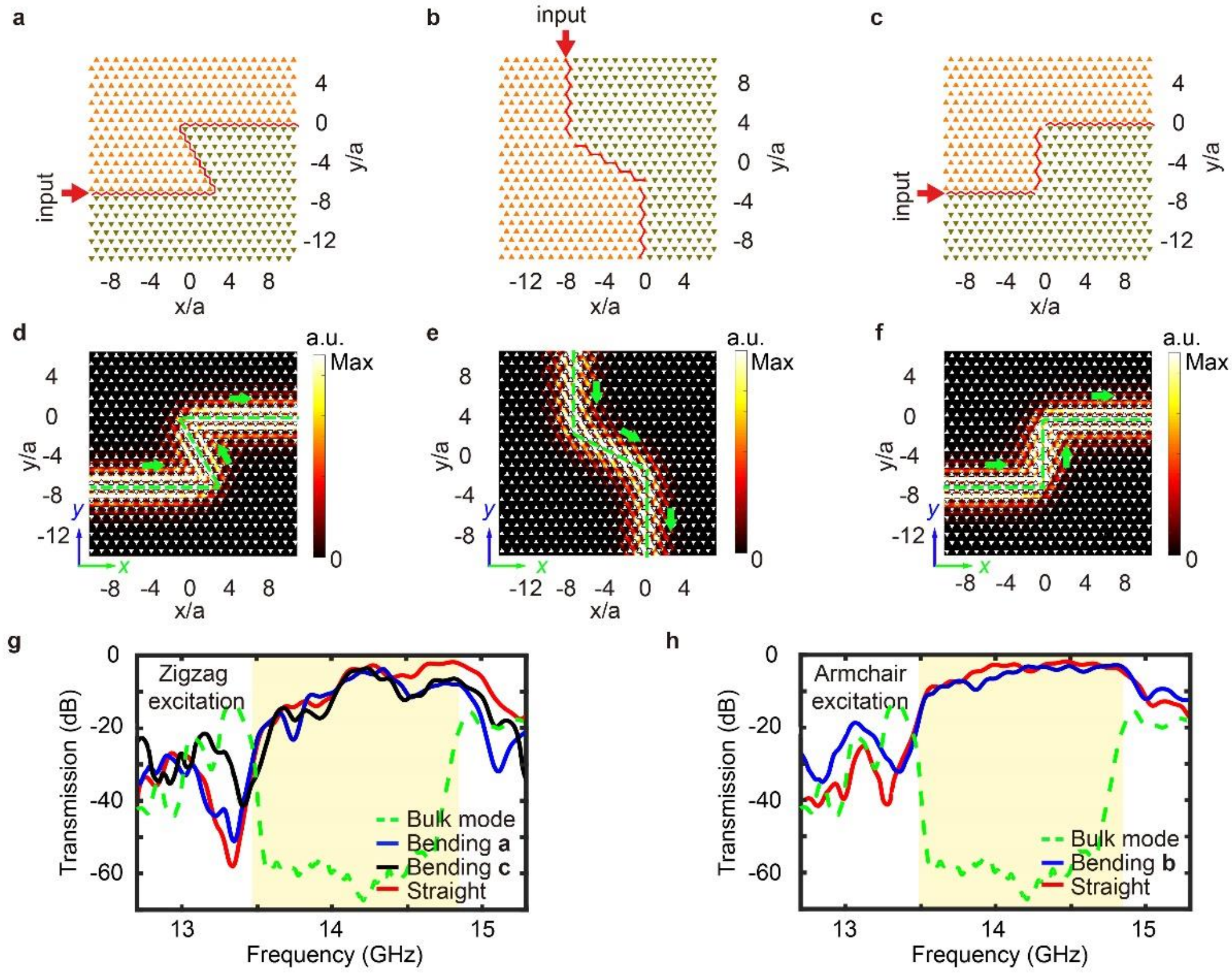

**Figure 3. Experimental demonstration of robust transport of valley kink states through sharp bending paths.** **a-c**, Scheme of sharply twisted topological interfaces. Twisted interfaces, with bending angles of 120°, 60°, and 90°, contain zigzag, armchair and a combination of zigzag and armchair interfaces, respectively. The red zigzag curves represent the interfaces. The orange/green triangles represent the scatterers with positive/negative rotation angles. **d-f**, Simulated electric field intensity distributions corresponding to **a-c**, respectively. Here, we place a point source near the left (d and f) or upper (e) terminations. The green curves represent the energy flows. **g, h**, Measured transmission for bending interfaces, straight interfaces (red curves) and bulk modes (green curves) when sources are placed at zigzag/armchair interfaces (marked as zigzag/armchair excitation). Blue and red curves in **g** represent transmissions of the valley kink states in **a, c**, respectively. The blue curve in **h** represents the transmission of the valley kink states in **b**. The yellow areas denote bandgaps.

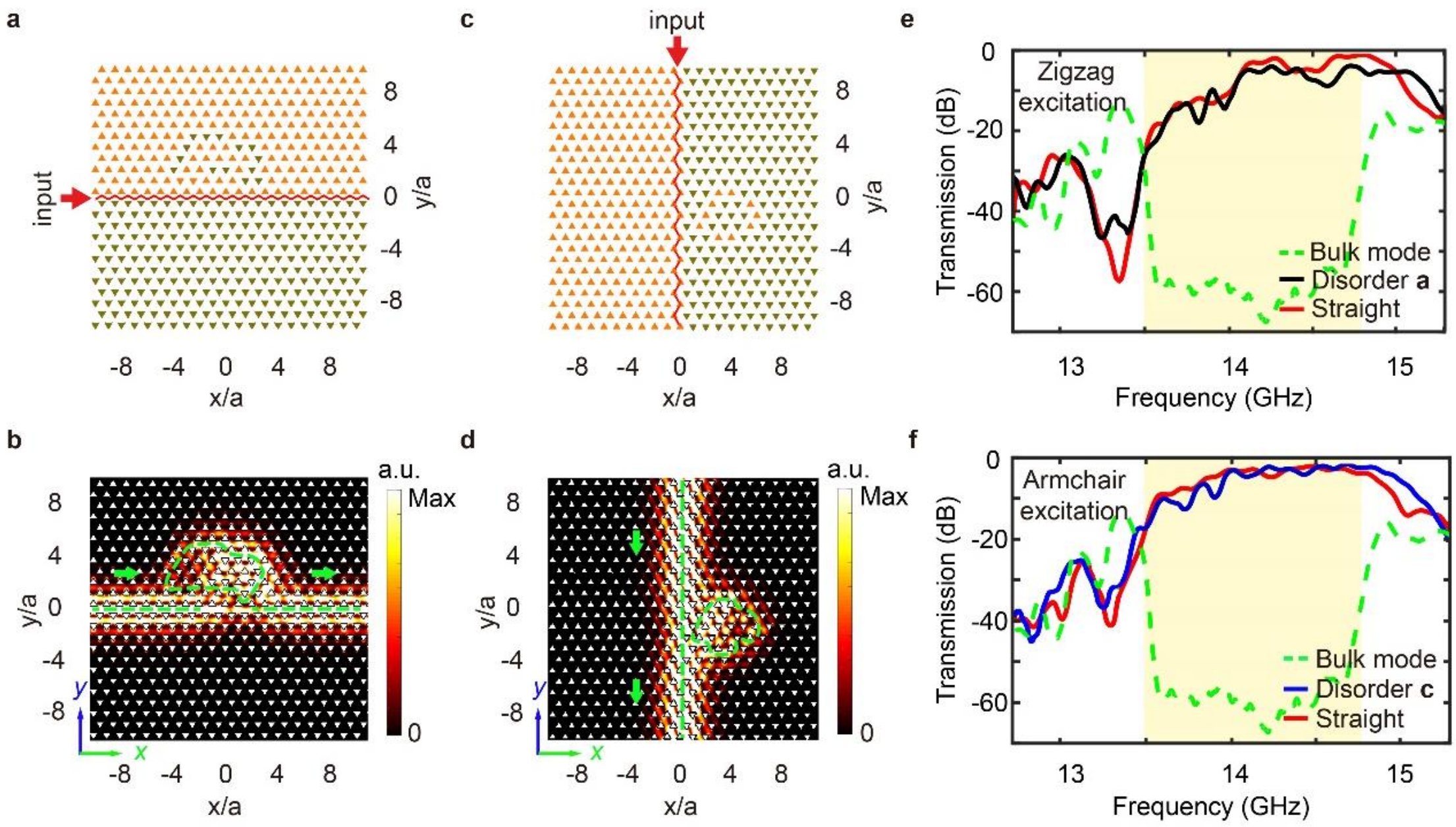

**Figure 4. Experimental demonstration of valley-polarized topological kink states through the interface with disorders. a, c**, Schematic view of zigzag and armchair interfaces with disorders. The red curves represent the interfaces. The orange/green triangles represent the scatterers with positive/negative rotation angles. **b, d**, Simulated $E_Z$ field intensity distributions corresponding to **a, c**. Here, we place a point source near the left (b) or upper (d) termination. The green curves represent energy flows. **e, f**, Measured transmission for the interfaces with disorders, straight interfaces (red curves) and bulk modes (green curves) when sources are placed at zigzag/armchair interfaces (marked as zigzag/armchair excitation). The black/blue curve represents the transmission of kink states in the structure corresponding to the schematic of **a/c**. The yellow regions represent the bandgaps.

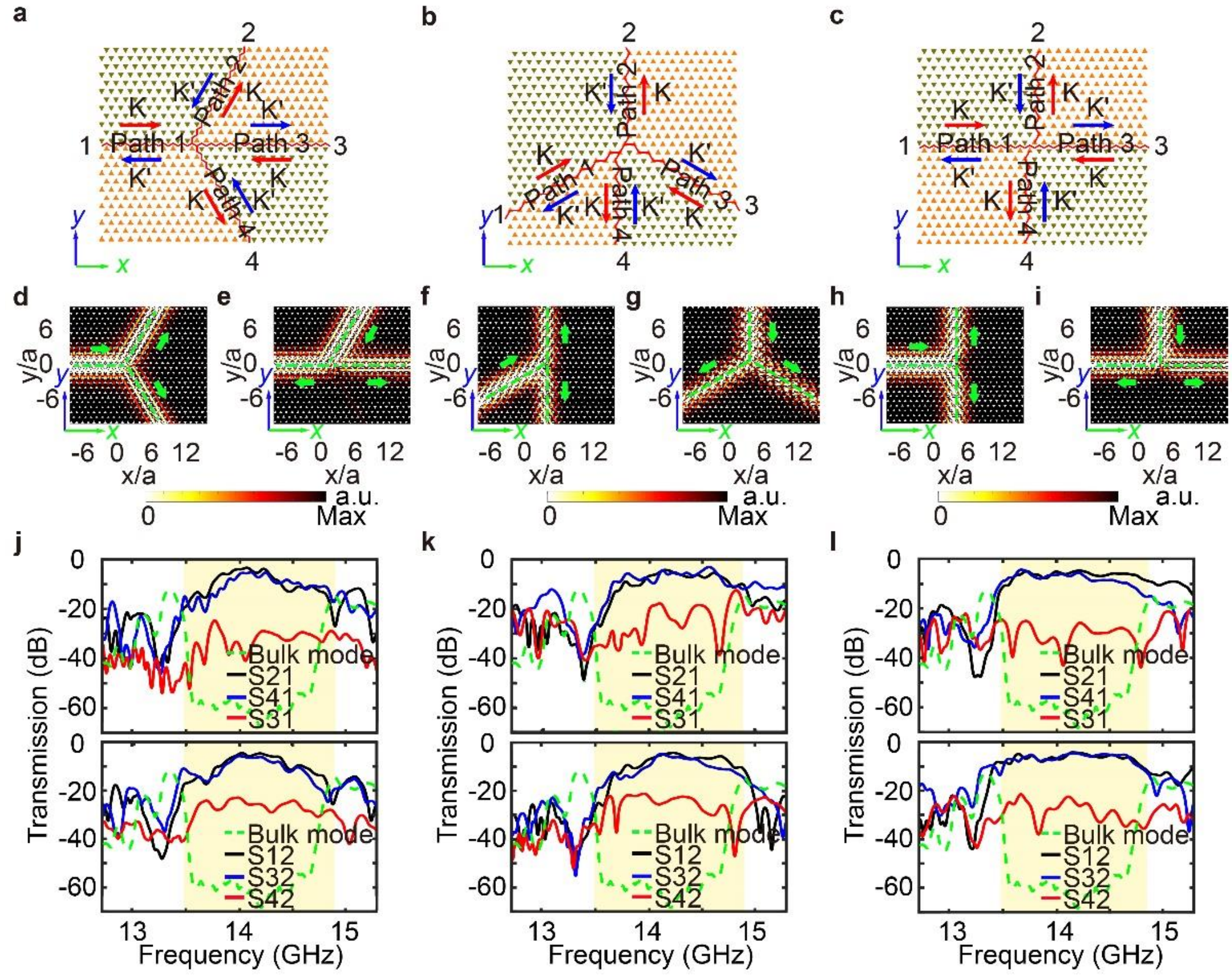

**Figure 5. Experimental demonstration of geometry-dependent topological channel intersections. a-c**, Schematic illustration of topological channel intersections. The orange/green triangles represent the scatterers with positive/negative rotation angles. The red and blue arrows represent propagating directions of kink states carrying K and K' valleys, respectively. **d-i**, Simulated $E_Z$ field intensity distributions when placing a point source at port 1 and port 2, respectively. The green curves represent the energy flows. **j-l**, Measured normalized transmissions when placing the source antenna at port 1 and port 2, respectively. The solid curves represent the transmissions of the valley kink states, and the dashed curve denotes the bulk transmission. The yellow region represents the bandgaps.